\documentclass[aip,jap,numerical,amssymb,
preprint,
reprint,
unsortedaddress
]{revtex4-1}

\usepackage{graphicx}
\usepackage{amsmath}

\begin{document}

\title{Heterointerface potentials in the effective-mass approximation for wurtzite semiconductor structures}

\author{Eduard Takhtamirov}
\affiliation{Conestoga College Institute of Technology and Advanced Learning, Kitchener, Ontario N2G 4M4, Canada}
\date{\today}

\begin{abstract}
In the effective-mass approximation, the step-like crystal potential of a wurtzite semiconductor heterostucture should be supplemented by Dirac delta-function heterointerface terms. They stem from the difference in the Bloch functions of the semiconductors and remain finite even for structures with graded chemical composition, where the terms are presented by a smeared Dirac delta function. We find these heterointerface potentials by employing the $k\cdot p$ method, and evaluate their strength from band-structure parameters of bulk materials. These potentials are weak for semiconductors compliant with the cubic approximation, which forces the zinc-blende crystal symmetry upon the wurtzite lattice. Nevertheless, they can produce a noteworthy effect due to a strong built-in electric field usually present in wurtzite heterostructures. We estimate that for GaN/AlN [0001] heterojunctions their net contributions to the energy of conduction and valence band states are 3~meV and 10~meV, respectively. The presence of the interface potential can modify the shape of the valence-band spectrum calculated without the potential. 
\end{abstract}

\maketitle

\section{Introduction}\label{intro}

In the last two decades, much effort has been devoted to accurate determination of the band offsets and polarization fields in group III-nitride semiconductor heterostructures. \cite{Murayama,Martin,Waldrop,Wei,VanDeWalle,Bernardini,Binggeli,Cociorva, Vurgaftman,Kuokstis,Westmeyer,Wu-offset,Cui,Shieh,Gladysiewicz} These material characteristics along with parameters of the bulk semiconductors are crucial elements of the band structure engineering of modern electronic and optoelectronic devices. \cite{Majewski,polar-effects,Wu} Meanwhile, it is tacitly supposed that the knowledge of the band offsets and polarization fields provides rather accurate description of the effective potential of a heterojunction. Can the simple model of a step-like crystal potential be defective?

\begin{figure}[tb]
\includegraphics [width=6cm]{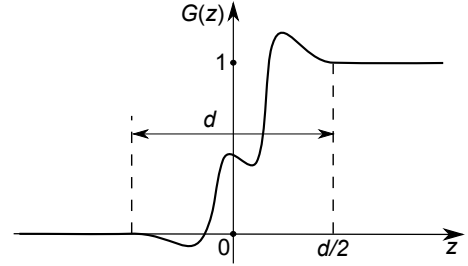}
\caption{Phenomenological one-dimensional form-factor $G(z)$ of a single heterojunction. The transition region of the junction is $d$, so that $G(z)\vert _{z < -d/2} = 0$ and  $G(z)\vert _{z > d/2} = 1$. }
\label{form-factor}
\end{figure}

Zhu and Kroemer were apparently the first to give a detailed analysis of the potential of an abrupt heterojunction to be used in the effective-mass (EM) equations. \cite{Kroemer,Zhu} They showed that the heterojunction can be described by a potential step to represent the band offset and a heterointerface potential in the form of a Dirac delta function, the strength of which is subject to the details of the heterointerface on an atomic scale. This result of the tight-binding model has later been confirmed by consistent application of the $k\cdot p$ method, \cite{Volkov,Takhtamirov} with the only input quantity, other than parameters of bulk materials, being a phenomenological coordinate-dependent form-factor $G(\mathbf r)$, see Fig.~\ref{form-factor}, that characterizes the chemical composition of a heterostructure. \cite{Leibler}

Contrary to what one might expect, the interface potential of a heterojunction is not formally represented by a space derivative of the function $G(\mathbf r)$. Briefly, its origin can be traced as follows. The form-factor can be decomposed into a gentle part, processed by using the standard technique of the EM approximation for slowly varying perturbations, \cite{Leibler,Luttinger} and a rapidly varying part, the space extent of which $d$ is supposed to be small on a scale of the wavelength $L$ of the envelope function (EF). The rapidly varying part can be processed in the spirit of the approach presented by Bir and Pikus \cite[Chap.~27]{Bir-Pikus} for the central-cell corrections in the problem of shallow impurity levels in bulk semiconductors. This rapidly varying part can accordingly be formulated as a convergent series of interface-localized generalized functions, \cite{Volkov,Takhtamirov} each successive term contributing less than the previous one in the parameter $a/L$, where $a$ is the lattice constant. It is the leading term of the series that gives rise to the delta-function potential of Zhu and Kroemer. The strength of the interface potentials not only depend on the microscopic details of the heterointerface but also on the orientation of the structure's growth direction with respect to the crystal axes. \cite{Takhtamirov_Nano}

Zhu and Kroemer pointed out that the heterointerface terms were small or negligible for GaAa/AlGaAs heterojunctions but might be strong in other semiconductor systems. \cite{Kroemer,Zhu} Since then, there have been obtained no experimental evidences of the existence of the diagonal in the band index interface potentials, which couple states of the same type. It is not surprising because even if they are present and properly included into the EM equations, the potentials do not modify symmetry but merely spectrum of the system, which may have been misinterpreted in terms of adjusted EM parameters or band offsets. On the other hand, the interface potentials can also produce interband couplings, which, though generally weaker than the intraband ones, are easily identified by unique effects they produce. Thus it has been shown for $[001]$-grown zinc-blende heterostructures that the interface terms are responsible for such phenomena as the intervalley $\Gamma$-$X_z$ \cite{Ando,Ivchenko_GX,gamma-x} and $X_x$-$X_y$ \cite{Ivchenko_GX} mixing of the conduction band states, mixing of light and heavy holes at the center of the two-dimensional Brillouin zone, \cite{Ivchenko_lh_hh_mixing} and also anisotropy of the interband optical transitions \cite{Krebs,Ivchenko_optical_anisotropy} and a twofold anisotropy of the in-plane spectrum of the conduction band electrons in asymmetric quantum wells. \cite{Takhtamirov_stripes,Reker}

Here we will ignore the rapidly varying part of the function $G(\mathbf r)$ to focus on another type of the interface potential, not considered so far, that is missing in zinc-blende heterostructures but present in wurtzite ones. It is proportional to a derivative of the form-factor $G(\mathbf r)$ and is thus finite even for structures with graded chemical composition where the rapidly varying part of the function $G(\mathbf r)$ produces an exponentially small contribution \cite[Chap.~22]{Bir-Pikus}.

\section{Heterointerface potentials in the $k\cdot p$ method}\label{body}

Let us consider a heterostructure composed of two semiconductors, so that the single-particle potential energy of an electron is
\begin{equation}
U=U (\mathbf r) = U_1+ G \delta U,
\end{equation}
where $U_1 = U_1\left( {\mathbf r}\right)$ is the periodic lattice potential of the nominally basis (potential well) material, while $\delta U = \delta U (\mathbf r) = U_2-U_1$ is the perturbative periodic potential and $U_2 = U_2( {\mathbf r})$ is the periodic potential of the barrier material. We suppose that the phenomenological function $G = G\left( {\mathbf r}\right)$ takes values of the order of unity or less even at the heterointerfaces. \cite{Leibler} Ideally, $G$ can even be a step-like function so that $G=0$ in the region of the potential well material and $G=1$ in the region of the barrier material. \cite{Takhtamirov} We also suppose that the semiconductors composing the structure are related, which means that $\delta U$ can be treated as a small perturbation as compared to the basis potential $U_1$.

We do not consider in details deformation effects appearing in lattice-mismatched pairs, which would lead to redundant complications, not essential for our results. We suppose that an inhomogeneous coordinate transformation restoring the unstrained lattice of the basis material and adjusting the lattice of the barrier material to that of the basis one has been performed, \cite{Zhang} so that in the transformed space both materials are described by the same lattice constant. The strain-induced piezoelectric field and the spontaneous polarization field are taken into account via the inclusion of an external scalar potential $W = W  \left( {\mathbf r}\right)$, while the strain modification of the band edge energies and the band offsets can be accounted for by using the strain Hamiltonian. \cite{Bir-Pikus,Chuang} For brevity, the latter as well as the spin-orbit interaction are considered only schematically. The strain-induced change in the matrix elements of the momentum operator $\mathbf p = \left( p_x,p_y,p_z \right)$ between Bloch states, being small in the parameter of the order of strain (several percent typically), is neglected.

Dealing with states near the Brillouin zone center, it is convenient to use the complete set of Kohn-Luttinger functions, \cite{Luttinger} with the unit-cell orthonormalized periodic parts $u_{n0} = u_{n0}\left( {\mathbf r}\right) \equiv \vert n \rangle$, which are chosen real, specified as follows:
\begin{equation}\label{w_basis}
\left( \frac{{\mathbf p}^2}{2m_0}+U_1 \right) u_{n0} =\epsilon _{n0} u_{n0},
\end{equation}
where $m_0$ is the free electron mass and $\epsilon _{n0}$ is the band edge energy of the $n$th band for the potential well material. We also introduce here zone-center Bloch functions $\widetilde u_{n0} = \widetilde u_{n0}\left( {\mathbf r}\right) \equiv \vert \widetilde n \rangle$ and the band edge energies $\widetilde \epsilon _{n0}$ for the barrier material:
\begin{equation}\label{b_basis}
\left( \frac{{\mathbf p}^2}{2m_0}+U_2 \right) \widetilde u_{n0} =\widetilde \epsilon _{n0} \widetilde u_{n0}.
\end{equation}
Treating $\delta U$ as a perturbation, \cite{Landavshitz} we have the following approximate expressions for the band edge energies:
\begin{equation}
\widetilde \epsilon _{n0} = \epsilon _{n0} + \delta U _{nn},
\end{equation}
and for the band edge Bloch functions:
\begin{equation}
\widetilde u_{n0} = u_{n0}+{\sum_{n'\ne n }}
\frac{ \delta U _{n'n} }
{\epsilon _{n0} - \epsilon _{n'0}}\,u_{n'0}, \label{left-right}
\end{equation}
where $\delta U _{n'n} = \langle n' \vert \delta U \vert n \rangle$.

Let us suppose that Cartesian coordinates are oriented so that the coordinate $z$ is along [0001] direction ($c$-axis) of wurtzite. Here we explicitly introduce the valence band Bloch functions $u_{\mathrm x 0} $ and $u_{\mathrm y 0} $, which transform as the coordinates $x$ and $y$, respectively, belonging to the representation $\Gamma_6$ of the space group $C_{6v}$, and also the valence band Bloch function $u_{\mathrm z 0}$ and the conduction band Bloch function $u_{\mathrm c 0}$, which transform as the coordinate $z$, belonging to the representation $\Gamma_1$. \cite{Bir-Pikus,Chuang}

In the mean field approximation, the Schr\"odinger equation reads: \cite{Landavshitz}
\begin{equation}\label{schroedinger}
\left( \frac{{\mathbf p}^2}{2m_0}+ U + W + H_{\mathrm {s}\varepsilon} \right) \Psi \left( {\mathbf r}\right)
=\epsilon \Psi \left( {\mathbf r}\right),
\end{equation}
with the term $H_{\mathrm {s}\varepsilon}$ comprising the spin-orbit interaction and the strain Hamiltonian. \cite{Bir-Pikus} Defining the $n$th band EF in $\mathbf r$-representation as $F_n = F_n\left( {\mathbf r}\right)$, we have the following expansion for the total wave function $\Psi \left( {\mathbf r}\right)$: \cite{Luttinger}
\begin{equation}\label{EF}
\Psi \left( {\mathbf r}\right) =\sum_n F_n \, u_{n0},
\end{equation}
where the summation is over all bands. The EFs satisfy the following $k\cdot p$ system of equations in the coordinate representation:
\begin{widetext}
\begin{equation}
\left( \epsilon _{n0} - \epsilon 
+ \frac{{\mathbf p}^2}{2m_0} + W\left( {\mathbf r} \right)
\right) F_n
+ \sum_{n^{\prime }}\left(
\frac{ {\mathbf p} \cdot \langle n \vert {\mathbf p} \vert n' \rangle }{m_0} 
+ G \left( {\mathbf r}\right) \delta U_{nn'} + H^{( \mathrm {s}\varepsilon )}_{nn'}
\right)
F_{n^\prime } = 0. \label{kp}
\end{equation}
\end{widetext}
Equation~(\ref{kp}) is obtained by treating the functions $G\left( {\mathbf r}\right)$ and $W\left( {\mathbf r}\right)$ as gentle. \cite{Leibler,Luttinger} At this point, we neglect all central-cell like corrections due to a rapid variation of the function $G\left( {\mathbf r}\right)$ at the heterointerfaces. \cite{Volkov,Takhtamirov} As in Eq.~(\ref{schroedinger}), the term $H^{( \mathrm {s}\varepsilon )}_{nn'} = H^{( \mathrm {s}\varepsilon )}_{nn'} (\mathbf r)$ takes into account contributions from the spin-orbit interaction and the strain Hamiltonian, now acting on EFs. The system of Equations (\ref{kp}) is valid for slowly varying functions $F_n$. \cite{Takhtamirov,Bir-Pikus}

The conduction and valence band EM equations can be obtained from the set of Eq.~(\ref{kp}) by using the L\"owdin perturbation scheme \cite{Bir-Pikus,Lowdin} up to the second order. For the conduction band, neglecting the spin-orbit interaction, we have:
\begin{equation}\label{c-band}
\left( H_{\mathrm c} + H_{ \mathrm c \varepsilon} + \Delta U_{\mathrm c} \right) F_{\mathrm c} = \epsilon F_{\mathrm c},
\end{equation}
where $H_{\mathrm c}$ is the ordinary conduction band Hamiltonian for a wurtzite heterostructure. It reads as follows:
\begin{equation}\label{H_c}
H_{\mathrm c} = \epsilon _{{\mathrm c}0} + \frac{p^2_z}{2m_z} + \frac{ p^2_x + p^2_y}{2m_x} 
+ G \delta U_{\mathrm {cc}} + W,
\end{equation}
where the parameter $\delta U_{\mathrm {cc}}$ defines the conduction band offset without strain contributions, $m_z$ and $m_x$ are the conduction band EMs,
\begin{equation}
\frac 1 {m_j} = \frac 1 {m_0} + \frac 2 {m^2_0}\sum_{n \ne \mathrm c}
\frac {\vert \langle \mathrm c \vert p_j \vert n \rangle \vert^2} {\epsilon_{{\mathrm c}0} - \epsilon _{n0}},
\end{equation}
with $j=z,x$, and $H_{ \mathrm c \varepsilon}$ is the conduction band strain Hamiltonian, \cite{Bir-Pikus,Chuang}
\begin{equation}
H_{ \mathrm c \varepsilon} = a_1 \varepsilon_{zz} + a_2 \left( \varepsilon_{xx} + \varepsilon_{yy}\right),
\end{equation}
where $a_1 = a_1 (\mathbf r)$ and $a_2 = a_2 (\mathbf r)$ are the material-dependent conduction band deformation potentials, and $\pmb \varepsilon = \pmb \varepsilon (\mathbf r)$ is the position-dependent strain tensor.

The term $\Delta U_{\mathrm c} = \Delta U_{\mathrm c} ({\mathbf r} )$ in the left-hand side of Eq.~(\ref{c-band}) is the new heterointerface potential introduced in this work,
\begin{equation}\label{U_c}
\Delta U_{\mathrm c} ({\mathbf r})= {\mathbf S} \cdot \left( {\pmb \nabla} G({\mathbf r}) \right),
\end{equation}
where the real vector $\mathbf S$ is
\begin{equation}\label{S}
\mathbf S = \frac \hbar {im_0} \sum_{n\ne \mathrm c} \frac {\langle \mathrm c \vert {\mathbf p} \vert n \rangle \delta U_{n{\mathrm c}}} {\epsilon_{{\mathrm c} 0} - \epsilon _{n0}}.
\end{equation}
It has been derived by Leibler, \cite{Leibler} but misinterpreted as producing a quasi-momentum shift in the position of the spectrum extrema, and thus concluded to vanish for high symmetry points of the Brillouin zone. Indeed, it takes place for $\Gamma$ point states in cubic crystals but not in wurtzite ones. It follows from the symmetry arguments, due to the fact that the operator $p_z$ belongs to the representation $\Gamma_1$ of the space group $C_{6v}$, \cite{Bir-Pikus} that the $S_z$ component of the vector $\mathbf S$ is still finite.

For the valence band, described by the band's EFs $F_\mathrm x$, $F_\mathrm y$ and $F_\mathrm z $, the system of the EM equations can be presented as follows:
\begin{equation}\label{valence}
\sum_{n'=\mathrm x, \mathrm y, \mathrm z} \left( H^{(\mathrm v)}_{nn'} + \Delta U_{nn'} \right) F_{n'} = \epsilon F_n, \quad n =\mathrm x, \mathrm y, \mathrm z.
\end{equation}
The matrix $\mathbf H^{(\mathrm v)}$ is the ordinary valence band Hamiltonian for wurtzite heterostructures, which includes the kinetic energy, step-like potential energy, spin-orbit interaction and strain Hamiltonians, \cite{Bir-Pikus,Chuang,Ren} given in Appendix~\ref{app} for reference. The matrix $\pmb \Delta {\mathbf U} = \pmb \Delta {\mathbf U} (\mathbf r)$ has the following finite components:
\begin{equation}
\begin{split}
&\Delta U_{\mathrm x \mathrm x} = \Delta U_{\mathrm y \mathrm y} = V_{\mathrm x \mathrm x} \frac {\partial G({\mathbf r})}{\partial z},\quad \Delta U_{\mathrm z \mathrm z} = V_{\mathrm z \mathrm z} \frac {\partial G({\mathbf r})}{\partial z},\\
&\Delta U_{\mathrm x \mathrm z} = -\delta N_3 G ({\mathbf r})p_x + V_{\mathrm x \mathrm z} \frac {\partial G({\mathbf r})}{\partial x},\\
&\Delta U_{\mathrm z \mathrm x} = \Delta U^\dag_{\mathrm x \mathrm z}= \delta N_3 G ({\mathbf r})p_x + V_{\mathrm z \mathrm x} \frac {\partial G({\mathbf r})}{\partial x},\\
&\Delta U_{\mathrm y \mathrm z} = -\delta N_3 G ({\mathbf r})p_y + V_{\mathrm x \mathrm z} \frac {\partial G({\mathbf r})}{\partial y},\\
&\Delta U_{\mathrm z \mathrm y} = \Delta U^\dag_{\mathrm y \mathrm z} = \delta N_3 G ({\mathbf r})p_y + V_{\mathrm z \mathrm x} \frac {\partial G({\mathbf r})}{\partial y}.
\end{split}
\end{equation}
The binary material parameters entering the above expressions are
\begin{equation}\label{V_j}
V_{jj} = \frac \hbar {im_0} \sum_{n\ne j} \frac {  \langle j \vert p_z \vert n \rangle \delta U_{n j}} {\epsilon_{j 0} - \epsilon _{n0}}, \quad j=\mathrm x, \mathrm z,
\end{equation}
%%%%%%%%
\begin{equation}
\begin{split}
\delta N_3 = \sum_{n \ne \mathrm x, \mathrm z}
&\left( \frac 1 {\epsilon_{\mathrm x 0} - \epsilon _{n0}} + \frac 1 {\epsilon_{\mathrm z 0} - \epsilon _{n0}} \right)\\
&\times
\frac {
\langle \mathrm z \vert p_x \vert n \rangle \delta U_{n \mathrm x} + \delta U_{\mathrm z n}
\langle n \vert p_x \vert \mathrm x \rangle }
{2m_0},
\end{split}
\end{equation}
\begin{equation}\label{V_xz}
\begin{split}
V_{\mathrm x \mathrm z} =
\sum_{n \ne \mathrm x, \mathrm z}
&\left( \frac 1 {\epsilon_{\mathrm x 0} - \epsilon _{n0}} + \frac 1 {\epsilon_{\mathrm z 0} - \epsilon _{n0}} \right)\\
&\times \frac { \hbar \langle \mathrm x \vert p_x \vert n \rangle \delta U_{n \mathrm z}}{2im_0},
\end{split}
\end{equation}
\begin{equation}\label{V_zx}
\begin{split}
V_{\mathrm z \mathrm x} =
\sum_{n \ne \mathrm x, \mathrm z}
&\left( \frac 1 {\epsilon_{\mathrm x 0} - \epsilon _{n0}} + \frac 1 {\epsilon_{\mathrm z 0} - \epsilon _{n0}} \right)\\
&\times \frac { \hbar \langle \mathrm z \vert p_x \vert n \rangle \delta U_{n \mathrm x}}{2im_0}.
\end{split}
\end{equation}

The matrix $\pmb \Delta {\mathbf U}$ has contributions of two types. The terms proportional to the parameter $\delta N_3$ provide position dependence for the linear in the momentum operator elements entering the kinetic energy matrix of Eq.~(\ref{Hk}), which are proportional to $N_3$, see Appendix~\ref{app}. The rest contributions, proportional to first partial derivatives of the form-factor $G({\mathbf r})$, are finite only at heterointerfaces.

\section{Estimates}

To estimate the above binary parameters given by Eqs.~(\ref{S}), (\ref{V_j}), (\ref{V_xz}), and (\ref{V_zx}), we restrict ourselves with the Kane model, \cite{Chuang,Kane} considering only conduction and three valence bands. However, it should be borne in mind that the estimation may lack accuracy for actual wide band gap materials such as GaN and AlN. We further neglect the energy $\Delta_{\mathrm {CR}}$ of the crystal splitting: \cite{Bir-Pikus, Chuang} $\epsilon_{\mathrm x 0} =\epsilon_{\mathrm y 0}= \epsilon_{\mathrm z 0} + \Delta_{\mathrm {CR}} \approx \epsilon_{\mathrm z 0}$, and introduce the band gap energy $E_g = \epsilon_{\mathrm c 0} - \epsilon_{\mathrm z 0}$ and the Kane matrix elements $P_1 = -i \hbar \langle \mathrm c \vert p_z \vert \mathrm z \rangle /m_0$ and $P_2 = -i \hbar \langle \mathrm c \vert p_x \vert \mathrm x \rangle /m_0$. With these notations and the adopted approximation we have:
\begin{equation}
\begin{split}
&S_z = V_{\mathrm z \mathrm z} = \frac {P_1 \delta U_{\mathrm c \mathrm z}}{E_g}, \quad 
V_{\mathrm x \mathrm z}= \frac {P_2 \delta U_{\mathrm c \mathrm z}}{E_g},\\
& V_{\mathrm x \mathrm x} = V_{\mathrm z \mathrm x}=0.
\end{split}
\end{equation}
To be consistent with the approximation of zero crystal splitting, we may use $P_2 \approx P_1$ so that all interface potentials are governed by a unique constant $\alpha $:
\begin{equation}\label{alpha}
\alpha = \frac {P_1 \delta U_{\mathrm c \mathrm z}} {E_g}.
\end{equation}
The conduction band interface potential of Eq.~(\ref{U_c}) is thus:
\begin{equation}
\Delta U_{\mathrm c} ({\mathbf r}) = \alpha \frac {\partial G({\mathbf r}) }{\partial z},
\end{equation}
while the valence band matrix $\pmb \Delta {\mathbf U} $ is
\begin{widetext}
\begin{equation}
\pmb \Delta {\mathbf U} =
\begin{pmatrix}
0 &0&
-\delta N_3 G ({\mathbf r})p_x + \alpha \frac {\partial G({\mathbf r})}{\partial x}
\cr
0&0&
-\delta N_3 G ({\mathbf r})p_y + \alpha \frac {\partial G({\mathbf r})}{\partial y}
\cr
\delta N_3 G ({\mathbf r})p_x + \alpha \frac {\partial G({\mathbf r})}{\partial x}
& \delta N_3 G ({\mathbf r})p_y +\alpha \frac {\partial G({\mathbf r})}{\partial y}
& \alpha \frac {\partial G({\mathbf r})}{\partial z}
\end{pmatrix}.\label{dU}
\end{equation}
\end{widetext}

Let us now demonstrate how knowing only bulk parameters of the materials composing the structure it is possible to extract the non-diagonal band offset $\delta U_{\mathrm c \mathrm z}$ and evaluate the parameter $\alpha$. We recall the following explicit expression for the parameter $A_7$ entering the kinetic energy Hamiltonian of Eq.~(\ref{Hk}), see Appendix~\ref{app}:
\begin{equation}\label{A7_w}
A_7 = \frac {\hbar \langle \mathrm z \vert p_x \vert \mathrm x \rangle} {i \sqrt 2 m_0}
\end{equation}
for the potential well material, and 
\begin{equation}\label{A7_b}
\widetilde A_7 = \frac {\hbar \langle \widetilde {\mathrm z} \vert p_x \vert \widetilde {\mathrm x} \rangle} {i \sqrt 2 m_0}
\end{equation}
for the barrier material. Using Eq.~(\ref{left-right}), adapted to the Kane model, we have:
\begin{equation}\label{left-right_Kane}
\langle \widetilde {\mathrm z} \vert \approx \langle \mathrm z \vert -
\frac {\delta U_{\mathrm c \mathrm z}}{E_g} \langle \mathrm c \vert,
\quad \vert \widetilde {\mathrm x} \rangle \approx \vert \mathrm x \rangle.
\end{equation}
From Eqs.~(\ref{A7_w}), (\ref{A7_b}) and (\ref{left-right_Kane}) we immediately obtain:
\begin{equation}
\delta U_{\mathrm c \mathrm z} = \frac{ \sqrt 2 E_g \left(A_7 - \widetilde A_7 \right)}{P_2},
\end{equation}
and, considering $P_2 \approx P_1$, we finally have:
\begin{equation}
\alpha = \sqrt 2 \left(A_7 - \widetilde A_7 \right).
\end{equation}

The strength of the interface potentials is very weak for systems based on wurtzite materials compliant with the cubic approximation. \cite{Bir-Pikus} For such semiconductors, in particular $\vert P_1 \vert \approx \vert P_2 \vert \gg \vert A_7 \vert$. For example, using the parameters of GaN, \cite{Vurgaftman} $m_z = 0.2 m_0$ and $E_g = 3.5$~eV, we have $P_1 \approx 8$~eV\AA, while $A_7 \approx 94$~meV\AA. \cite{Vurgaftman,Ren} Vurgaftman and Meyer \cite{Vurgaftman} report $A_7 =0$ for wurtzite AlN, so that for GaN/AlN heterostructures $\alpha \approx 130$~meV\AA, while the interband offset $\delta U_{\mathrm c \mathrm z}$ is as small as 60~meV. The latter estimate has already been reported. \cite{takhtamirov_jap} Under these conditions, the inclusion of the interface potentials in the EM Hamiltonians cannot modify the states essentially. In particular, one may not expect that they can produce Tamm-like states localized at heterointerfaces.

\begin{figure}[t]
\includegraphics [width=6cm]{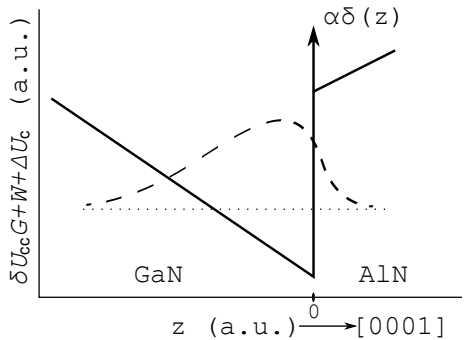}
\caption{Potential energy profile of the conduction band for a heterojunction GaN/AlN, solid line, including the interface potential, arrowed line. Also shown are the ground eigenenergy, dotted line, and the corresponding envelope function, dashed line.}
\label{quantum-well}
\end{figure}

Let us evaluate the effect of the interface potential on the spectrum of the conduction band states in a quantum well formed by a single heterojunction GaN/AlN in the model of the mathematically abrupt form-factor $G(z) = \Theta (z)$, where $\Theta (z)$ is the Heaviside step function, so that $d \Theta (z)/d z = \delta (z)$, see Fig.~\ref{quantum-well}, where $\delta (z)$ is the Dirac delta function.
Neglecting for brevity the strain Hamiltonian $H_{ \mathrm c \varepsilon}$ in Eq.~(\ref{c-band}), we consider the Hamiltonian $H^{(\mathrm {tot})}_{\mathrm c} = H_{\mathrm c} + \Delta U_{\mathrm c}$:
\begin{equation}\label{H_EMA}
H^{(\mathrm {tot})}_{\mathrm c}  = \epsilon _{{\mathrm c}0}  + \frac{p^2_z}{2m_z} + \frac{ p^2_x + p^2_y}{2m_x} 
+ \Theta(z) \delta U_{\mathrm {cc}} + W(z) + \alpha \delta(z).
\end{equation}
As soon as the term $\Delta U_{\mathrm c} = \alpha \delta(z)$ is a small perturbation, the discrete spectrum of the Hamiltonian of Eq.~(\ref{H_EMA}) can be approximated as follows:
\begin{equation}
\epsilon_N = \epsilon^{(0)}_N + \alpha ( N\vert \delta(z) \vert N),
\end{equation} 
where $\epsilon^{(0)}_N$ are eigenenergies of the Hamiltonian $H_{\mathrm c}$ with the corresponding EFs $F^{(0)}_N = \vert N )$, that is $H_{ \mathrm c}\vert N ) =  \epsilon^{(0)}_N \vert N )$.

Diagonal matrix elements of the commutator $[p_z , H_{ \mathrm c}]$ disappear for any two-dimensional subband index $N$:
\begin{equation}\label{comm}
( N\vert p_z H_{ \mathrm c} - H_{ \mathrm c} p_z \vert N)=0,
\end{equation}
resulting in the following identity: \cite{Bir-Pikus}
\begin{equation}\label{vp-zero}
\delta U_{\mathrm {cc}} ( N\vert \delta(z) \vert N) = - ( N\vert \frac {dW(z)}{dz} \vert N).
\end{equation}
The right-hand side of Eq.~(\ref{vp-zero}) is proportional to the average build-in electric field mostly originating due to the spontaneous polarization. \cite{Cui} If we neglect the contribution of the AlN barrier region to the matrix element on the right-hand side of Eq.~(\ref{vp-zero}), we will have: $( N\vert \frac {dW(z)}{dz} \vert N)\approx -e E_{\mathrm {GaN}}$, where $-e<0$ is the electron charge and $E_{\mathrm {GaN}}$ is the polarization field in the quantum well material (GaN). If we use $\delta U_{\mathrm {cc}} = 2.5$~eV, which is accepted as the conduction band offset for a heterojunction formed by AlN lattice matched to GaN, \cite{Bernardini,Vurgaftman} and $E_{\mathrm {GaN}}=5$~MV/cm, \cite{Cui} we have the following estimate: $( N\vert \delta(z) \vert N) \approx 0.02$~\AA$^{-1}$, and $ \alpha ( N\vert \delta(z) \vert N) \approx 3$~meV for the values of the parameters $A_7$ given by Vurgaftman and Meyer. \cite{Vurgaftman}

For the valence band, the above arguments expressed by Eqs.~(\ref{comm}) and (\ref{vp-zero}) are not directly applicable because the interface potential is not purely diagonal. As usual, non-diagonal components of the potential make a minor contribution, but the potential still has a diagonal component to influence the $z$-component of the EF. Consequently, when the interface potential is included, the spectrum of the hole states may not be uniformly shifted as we observed in the conduction-band case, but the in-plane effective masses of the subbands will also be modified. The interface potential has the strongest influence on the states with nearly isotropic in-plane electron density. For such states, the conduction-band estimation approach can be used. Then, for a heterojunction formed by GaN lattice matched to AlN, we must take into account smaller band offset $\delta U_{\mathrm {zz}} = 0.85$~eV \cite{Bernardini,Vurgaftman}, which results in the contribution of the interface potentials of the order of 10~meV.

Note that besides the restrictions of the four-band Kane model, applicability of which to wide band gap semiconductors is questionable, the values of the bulk parameters $A_7$ cannot be considered as established in the literature, so that the above figure for $\alpha$ may thus be unreliable. For example, using the parameters given by Rinke {\em et al.}, \cite{Rinke} $A_7 \approx 46$~meV\AA~(GaN) and $\widetilde A_7 \approx 26$~meV\AA~(AlN), we derive $\alpha \approx 30$~meV\AA, which is four times as small as the figure based on the parameters given by Vurgaftman and Meyer. \cite{Vurgaftman}

\section{Conclusions}

Using the $k\cdot p$ method, we have shown that the potential of a wurtzite semiconductor heterostructure must be supplemented by Dirac delta-function terms located at heterointerfaces. These heterointerface potentials are present even for structures with graded chemical composition, and their strength is governed by binary material parameters that can be deduced knowing only band structure characteristics of the bulk semiconductors. They are weak for materials whose wurtzite lattice slightly deviates from that of a cubic crystal. We have obtained these potentials for electrons and holes and estimated their strength. In the four-band Kane model, the potentials are defined by a unique parameter proportional to the difference in the parameters $A_7$ of the semiconductors of the structure. For GaN/AlN [0001] heterojunctions, we have evaluated that the contribution of the potentials to the energy of conduction and valence band states is 3~meV and 10~meV, respectively. In addition, the presence of the interface potential can modify the shape of the valence-band spectrum. 

\section*{Acknowledgments}

A part of this work has been done at Wilfrid Laurier University, Ontario, Canada.

\appendix

\section{Valence band Hamiltonian}\label{app}

In the basis of the band edge Bloch functions $u_{\mathrm x}$, $u_{\mathrm y}$ and $u_{\mathrm z}$, the valence band Hamiltonian in the EM approximation without interface potentials is as follows:
\begin{equation}\label{H}
\mathbf H^{(\mathrm v)} = \mathbf H^{(0)} + \mathbf H^{(\varepsilon)} + \mathbf H^{(\sigma)} + \mathbf H^{(k)},
\end{equation}
where $\mathbf H^{(0)}$ is the position-dependent potential energy of an electron:
\begin{widetext}
\begin{equation}
\mathbf H^{(0)} =
\begin{pmatrix}
\epsilon_{\mathrm x 0} + G \delta U_{\mathrm x \mathrm x} +W &0&0\cr
0&\epsilon_{\mathrm y 0} + G \delta U_{\mathrm y \mathrm y} +W&0\cr
0&0&\epsilon_{\mathrm z 0} + G \delta U_{\mathrm z \mathrm z} +W
\end{pmatrix},\label{H0}
\end{equation}
where $\epsilon_{\mathrm x 0} = \epsilon_{\mathrm y 0}$, while $\delta U_{\mathrm x \mathrm x} = \delta U_{\mathrm y \mathrm y}$ and $\delta U_{\mathrm z \mathrm z}$ are the valence band offsets, not including stain effects, which appear in the strain Hamiltonian
\begin{equation}\label{He}
\mathbf H^{(\varepsilon)} = \begin{pmatrix}
l_1 \varepsilon_{xx} + m_1 \varepsilon_{yy} + m_2 \varepsilon_{zz} &
n_1 \varepsilon_{xy} & n_2 \varepsilon_{xz}\cr
n_1 \varepsilon_{xy} &
m_1 \varepsilon_{xx} + l_1 \varepsilon_{yy} + m_2 \varepsilon_{zz} &
n_2 \varepsilon_{yz}\cr
n_2 \varepsilon_{xz} & n_2 \varepsilon_{yz} &
m_3\left( \varepsilon_{xx} + \varepsilon_{yy}\right) + l_2 \varepsilon_{zz}
\end{pmatrix},
\end{equation}
\end{widetext}
where real material-dependent parameters $l_1$, $l_2$, $m_1$, $m_2$, $n_1$, and $n_2$ are expressed via conventional \cite{Bir-Pikus,Chuang,Ren,Vurgaftman} components of the deformation potential tensor as follows:
\begin{equation}
\begin{split} \label{lmn_thru_D}
&l_1 = D_2 +D_4 +D_5, \quad m_1 = D_2 +D_4 -D_5,\\
&n_1 = 2 D_5,\quad l_2 = D_1, \quad m_2 = D_1+D_3,\\
&n_2 = \sqrt 2 D_6, \quad m_3 = D_2.
\end{split}
\end{equation}

The Hamiltonian $\mathbf H^{(\sigma)}$ in Eq.~(\ref{H}) describes the spin-orbit interaction: \cite{Bir-Pikus,Chuang}
\begin{eqnarray} \label{Hso}
\mathbf H^{(\sigma)}= 
\begin{pmatrix}
0&-i\Delta_2 \sigma_z&i\Delta_3 \sigma_y\cr
i\Delta_2 \sigma_z&0&-i\Delta_3 \sigma_x\cr
-i\Delta_3 \sigma_y&i\Delta_3 \sigma_x&0
\end{pmatrix},
\end{eqnarray}
\newline
where $\sigma_x$, $\sigma_y$, and $\sigma_z$ are the Pauli matrices, \cite{Landavshitz}
and real $\Delta_2 = \Delta_2\left(\mathbf r \right)$ and $\Delta_3= \Delta_3\left(\mathbf r \right)$ are the parameters of the valence-band spin-orbit splitting. \cite{Bir-Pikus,Chuang} The valence band EFs $F_\mathrm x$, $F_\mathrm y$, and $F_\mathrm z $ are thus defined as two-component spinors.

The kinetic energy Hamiltonian $\mathbf H^{(k)}$ of Eq.~(\ref{H}) is
\begin{widetext}
\begin{equation}\label{Hk}
\mathbf H^{(k)} = \begin{pmatrix}
L_1 p^2_x + M_1 p^2_y + M_2 p^2_z & N_1 p_x p_y & N_2 p_x p_z - N_3 p_x\cr
N_1 p_x p_y & M_1 p^2_x + L_1 p^2_y + M_2 p^2_z & N_2 p_y p_z - N_3 p_y\cr
N_2 p_x p_z+N_3 p_x & N_2 p_y p_z +N_3 p_y& M_3\left( p^2_x + p^2_y \right) + L_2 p^2_z
\end{pmatrix},
\end{equation}
\end{widetext}
where
\begin{equation}
\begin{split} \label{LMN_thru_A1}
&L_1 = \frac {1}{2m_0}\left( A_2 +A_4 +A_5 \right),
\quad L_2 = \frac {1}{2m_0}A_1, \\
&M_1 = \frac {1}{2m_0}\left( A_2 +A_4 -A_5\right),
\quad N_1 = \frac {1}{2m_0} 2 A_5,
\end{split}
\end{equation}
% split 4 lines into 2x2 lines just to make it more flexible on the page,
%no refs on the equations
\begin{equation}
\begin{split} \label{LMN_thru_A2}
&M_2 = \frac {1}{2m_0}\left( A_1+A_3 \right),
\quad N_2 = \frac {1}{2m_0} \sqrt 2 A_6\\
&M_3 = \frac {1}{2m_0} A_2, \quad N_3 = \frac i \hbar \sqrt 2 A_7,
\end{split}
\end{equation}
and $A_1$, $A_2$, \ldots $A_7$ are real material parameters in conventional notations. \cite{Bir-Pikus,Chuang,Ren,Vurgaftman}

If it is preferable to have the valence band EM equations expressed in terms of the EFs $F_1$, $F_2$, and $F_3$ that multiply the conventional \cite{Bir-Pikus,Chuang} basis functions $u_1 = (u_{\mathrm x}+iu_{\mathrm y})/\sqrt 2$, $u_2 = (u_{\mathrm x} - iu_{\mathrm y})/\sqrt 2$, and $u_3 = i u_{\mathrm z}$, one can use the canonical transformation
\begin{equation}
\begin{pmatrix}
F_{\mathrm x}\\
F_{\mathrm y}\\
F_{\mathrm z}
\end{pmatrix}
=
\begin{pmatrix}
\frac 1 {\sqrt 2}&\frac 1 {\sqrt 2}&0\\
\frac i {\sqrt 2}&-\frac i {\sqrt 2}&0\\
0&0&i
\end{pmatrix}
\begin{pmatrix}
F_1\\
F_2\\
F_3
\end{pmatrix}
\equiv \mathbf T
\begin{pmatrix}
F_1\\
F_2\\
F_3
\end{pmatrix},
\end{equation}
so that Eq.~(\ref{valence}) now reads
\begin{equation}\label{valence_123}
\sum_{n'=1, 2, 3} \left( \hat H^{(\mathrm v)}_{nn'} + {\Delta  \hat U}_{nn'} \right) F_{n'} = \epsilon F_n, \quad n =1, 2, 3,
\end{equation}
where $\hat {\mathbf H}^{(\mathrm v)} = {\mathbf T}^{-1}{\mathbf H}^{(\mathrm v)}{\mathbf T}$ and  
${\pmb \Delta \hat {\mathbf U}} = {\mathbf T}^{-1} {\pmb \Delta \mathbf U} {\mathbf T}$.

\end{document}